\documentclass[12pt]{article}
\usepackage{amsmath}
\usepackage[unicode,bookmarks,bookmarksopen,bookmarksopenlevel=2,colorlinks,linkcolor=blue,citecolor=green]{hyperref}

\input{tcilatex}
\begin{document}

\title{Classification of conservative hydrodynamic chains. Vlasov type
kinetic equation, Riemann mapping and the method of symmetric hydrodynamic
reductions.}
\author{Maxim V. Pavlov$^{1}$, Sergei Zykov$^{2,3}$ \\
$^{1}$Department of Mathematical Physics,\\
P.N. Lebedev Physical Institute of Russian Academy of Sciences,\\
Moscow, Leninskij Prospekt, 53\\
$^{2}$Department of Mathematical Physics,\\
SISSA, Trieste, Italy\\
$^{3}$Institute of Metal Physics, Ural branch of RAS,\\
Ekaterinburg, Russia}
\date{}
\maketitle

\begin{abstract}
A complete classification of integrable conservative hydrodynamic chains is
presented. These hydrodynamic chains are written via special coordinates --
moments, such that right hand sides of these infinite component systems
depend linearly on a discrete independent variable $k$. All variable
coefficients of these hydrodynamic chains can be expressed via modular forms
with respect to moment $A^{0}$, via hypergeometric functions with respect to
moment $A^{1}$; they depend polynomially on moment $A^{2}$ and linearly on
all other higher moments $A^{k}$. A dispersionless Lax representation is
found. Corresponding collisionless Boltzmann (Vlasov like kinetic) equation
is derived. A Riemann mapping is constructed. A generating function of
conservation laws and commuting flows is presented.
\end{abstract}

\tableofcontents

hydrodynamic chains, Riemann invariants, symmetric hydrodynamic type systems%
\textit{.}

\bigskip


\newpage

\section{Introduction}

In past years (2003 up to now) significant results were obtained in the
theory of integrable hydrodynamic chains (see \cite{MaksEps}, \cite{egor}, 
\cite{FerMarsh}, \cite{algebra}, \cite{maksbenney}). A first integrable
hydrodynamic chain%
\begin{equation}
A_{t}^{k}=A_{x}^{k+1}+kA^{k-1}A_{x}^{0}\text{, \ \ \ }k=0,1,2,...  \label{1}
\end{equation}%
was derived by D. Benney (see \cite{Benney}) in 1973. An integrability of
the Benney hydrodynamic chain can be illustrated by an existence of a
generating function of conservation laws (see \cite{Benney}, \cite{KM})%
\begin{equation}
p_{t}=\left( \frac{p^{2}}{2}+A^{0}\right) _{x},  \label{2}
\end{equation}%
where a generating function of conservation law densities is given by%
\begin{equation}
p=\lambda -\frac{H_{0}}{\lambda }-\frac{H_{1}}{\lambda ^{2}}-\frac{H_{2}}{%
\lambda ^{2}}-...,  \label{3}
\end{equation}%
whose all conservation law densities are polynomial functions with respect
to moments $A^{k}$, i.e. $%
H_{0}=A^{0},H_{1}=A^{1},H_{2}=A^{2}+(A^{0})^{2},H_{3}=A^{3}+A^{0}A^{1},...$
It means that hydrodynamic chain (\ref{1}) also can be written in the
conservative form (see, for instance, \cite{MaksGen})%
\begin{equation}
\partial _{t}H_{0}=\partial _{x}H_{1},\text{ \ \ }\partial _{t}H_{k}=\left(
H_{k+1}-\frac{1}{2}\underset{m=0}{\overset{k-1}{\sum }}H_{m}H_{k-1-m}\right)
_{x},\text{ \ }k=1,2,...  \label{4}
\end{equation}%
We are interested in a description of integrable hydrodynamic chains written
in the form (cf. (\ref{1}))%
\begin{equation}
A_{t}^{k}=f_{1}A_{x}^{k+1}+f_{0}A_{x}^{k}+A^{k+1}(s_{0}A_{x}^{0}+s_{1}A_{x}^{1})+A^{k}(r_{0}A_{x}^{0}+r_{1}A_{x}^{1})
\label{lin}
\end{equation}%
\begin{equation*}
+k[A^{k+1}(w_{0}A_{x}^{0}+w_{1}A_{x}^{1}+w_{2}A_{x}^{2})+A^{k}(v_{0}A_{x}^{0}+v_{1}A_{x}^{1}+v_{2}A_{x}^{2})+A^{k-1}(u_{0}A_{x}^{0}+u_{1}A_{x}^{1}+u_{2}A_{x}^{2})],
\end{equation*}%
where coefficients $f_{i},s_{j},r_{k}$ depend on first two moments $A^{0}$
and $A^{1}$ only, while all other coefficients $w_{m},v_{n},u_{p}$ depend
just on first three moments $A^{0},A^{1}$ and $A^{2}$.

Recently, two particular cases of hydrodynamic chains (\ref{lin}) were
completely investigated. The Hamiltonian\ hydrodynamic chains (here $\mathbf{%
H}_{1,k}\equiv \partial \mathbf{H}_{1}/\partial A^{k},k=0,1$) 
\begin{equation*}
A_{t}^{k}=(\alpha +\beta )\mathbf{H}_{1,1}A_{x}^{k+1}+\beta \mathbf{H}%
_{1,0}A_{x}^{k}+[\alpha (k+1)+2\beta ]A^{k+1}(\mathbf{H}_{1,1})_{x}+(\alpha
k+2\beta )A^{k}(\mathbf{H}_{1,0})_{x}
\end{equation*}%
are associated with the Kupershmidt Poisson brackets (see \cite{FKMP} and 
\cite{KuperNorm}); while the Hamiltonian\ hydrodynamic chains (here $\mathbf{%
H}_{2,k}\equiv \partial \mathbf{H}_{2}/\partial A^{k},k=0,1,2$)%
\begin{equation*}
A_{t}^{k}=2\mathbf{H}_{2,2}A_{x}^{k+1}+\mathbf{H}%
_{2,1}A_{x}^{k}+(k+2)A^{k+1}(\mathbf{H}_{2,2})_{x}+(k+1)A^{k}(\mathbf{H}%
_{2,1})_{x}+kA^{k-1}(\mathbf{H}_{2,0})_{x}
\end{equation*}%
are associated with the Kupershmidt--Manin Poisson bracket (see the second
part in \cite{FerMarsh}, \cite{GibRai} and \cite{KuperNorm}). These
hydrodynamic chains are integrable if and only if all components of
corresponding Haantjes tensors vanish. It means that the corresponding
Hamiltonian densities $\mathbf{H}_{1}(A^{0},A^{1})$ and $\mathbf{H}%
_{2}(A^{0},A^{1},A^{2})$ cannot be arbitrary. A full list of admissible
expressions is given in \cite{FKMP} and \cite{FerMarsh}, respectively.

In a general case, the coefficient $f_{1}$ in (\ref{lin}) is reducible to
the unity, the coefficient $f_{0}$ can be eliminate by an appropriate change
of moments $A^{k}$; while all other coefficients can be simplified in the
integrable case only. Following the approach based on an existence of first
three conservation laws and vanishing of the Haantjes tensor (see the first
part in \cite{FerMarsh}), one can extract the \textit{integrable} case%
\begin{equation}
A_{t}^{k}=A_{x}^{k+1}-k[(A^{k+1}+u_{0}A^{k}+u_{-1}A^{k-1})[\ln (A^{2}+\sigma
)]_{x}-A^{k}(u_{0})_{x}-A^{k-1}(u_{-1})_{x}],  \label{i}
\end{equation}%
where functions $u_{0},u_{-1},\sigma $ satisfy to an overdetermined system
in an involution (see (\ref{usi})). The same result can be obtained by the
method of hydrodynamic reductions established by J. Gibbons and S.P. Tsarev
in \cite{GT} and developed by E.V. Ferapontov and K.R. Khusnutdinova in \cite%
{FK}. In this paper, we utilize the concept of the so-called \textit{%
symmetric} hydrodynamic reductions (see \cite{algebra}, \cite{MaksHam}). In
this case, an existence of a Riemann mapping $\lambda
(q,A^{0},A^{1},A^{2},...)$ connecting the Vlasov type kinetic equation (see 
\cite{Gibbons}, \cite{OPS}, \cite{zakh}) with hydrodynamic chain (\ref{i})
leads to an overdetermined system in involution (\ref{full}), (\ref{zuk}), (%
\ref{zik}), (\ref{zak}), (\ref{zyk}), (\ref{sikma}), whose general solution
can be parameterized by hypergeometric functions. Then a generating function
of conservation laws can be found in quadratures. Thus, an infinite series
of conservation laws densities $H_{k}(A^{0},A^{1},...,A^{k})$ allows to
rewrite hydrodynamic chain (\ref{i}) in the conservative form\footnote{%
A problem of a description of integrable hydrodynamic chains (\ref{tri}) was
formulated in \cite{MaksEps}. A particular and important Egorov's case $%
F_{0}(H_{0},H_{1})\equiv H_{1}$ was investigated in \cite{egor}.} (cf. (\ref%
{4}))%
\begin{equation}
\partial _{t}H_{k}=\partial _{x}F_{k}(H_{0},H_{1},...,H_{k+1}),\text{ \ }%
k=0,1,2,...  \label{tri}
\end{equation}%
We \textbf{prove} that its first two conservation laws coincide with first
two conservation laws found in \cite{FerMarsh}. In the general case, E.V.
Ferapontov and D.G. Marshall found that $F_{2}(H_{0},H_{1},H_{2})=\ln
H_{2}+G(H_{0},H_{1})$ and functions $F_{0}(H_{0},H_{1}),G(H_{0},H_{1})$
satisfy to another overdetermined system in involution (see \cite{FerMarsh}%
). Moreover, we \textbf{prove} that this system in involution (\ref{ferma})
is equivalent to system in involution (\ref{full}), (\ref{zuk}), (\ref{zik}%
), (\ref{zak}), (\ref{zyk}), (\ref{sikma}). Thus, a complete classification
of \textit{integrable conservative hydrodynamic chains} (\ref{tri}) is given
in this paper.

This paper is organized in the following way. In Section 2, symmetric $2N$
component hydrodynamic reductions are extracted by virtue of Zakharov's
moment decomposition (see \cite{MaksHam}, \cite{zakh}). Such $2N$ component
hydrodynamic type systems contain $N$ component symmetric sub-systems, which
are still hydrodynamic reductions. These $N$ component hydrodynamic type
systems imply to the Vlasov type kinetic equation. We show that an existance
of the Riemann mapping connecting this Vlasov type kinetic equation with
hydrodynamic chain (\ref{i}) allows to select all integrable hydrodynamic
chains. In Section 3, canonical coordinates in a moment space are
introduced. Then a further investigation simplifies. In Section 4, a special
\textquotedblleft triangular\textquotedblright\ case is completely
integrated. Three variable coefficients in (\ref{i}) can be parameterized by
solutions of the so-called Halphen--Darboux system (see \cite{Abl}). In
Section 5, a generating function of conservation laws is found. In
Conclusion, a generalization of the approach presented in this paper is
discussed.

\section{Zakharov's moment decomposition}

The moment decomposition approach developed in \cite{MaksHam} (see also \cite%
{algebra}) is based on a concept of an existence of symmetric hydrodynamic
type systems%
\begin{equation*}
a_{t}^{i}=\partial _{x}F(\mathbf{a};p)|_{p=a^{i}}\text{, \ \ \ }i=1,2,...,N,
\end{equation*}%
which are nothing but hydrodynamic reductions of the hydrodynamic chains,
where in all known cases before (see, for instance, \cite{FKMP}, \cite%
{algebra}, \cite{MaksGen}, \cite{MaksHam}, \cite{maksbenney}, \cite%
{MaksKuper}) each corresponding moment $A^{k}$ depends on $N$ functions of a
single variable\footnote{%
let us emphasize that $N$ is an arbitrary natural number.}, i.e.%
\begin{equation*}
A^{k}=\sum_{m=0}^{N}f_{mk}(a^{m}),\text{ \ }k=1,2,3,...
\end{equation*}%
Another moment decomposition (introduced by V.E. Zakharov, see \cite{zakh}) 
\begin{equation}
A^{k}=\sum_{m=0}^{N}(a^{m})^{k}b^{m}  \label{zakh}
\end{equation}%
also is applicable in all these known cases (see \cite{MaksHam}). Benney
hydrodynamic chain (\ref{1}) under this moment decomposition reduces to the $%
2N$ component hydrodynamic type system%
\begin{equation*}
a_{t}^{i}=\left( \frac{(a^{i})^{2}}{2}+A^{0}\right) _{x},\text{\ \ }%
b_{t}^{i}=(a^{i}b^{i})_{x},
\end{equation*}%
which possesses a formal reduction to the $N$ component case (cf. (\ref{2}))%
\begin{equation*}
a_{t}^{i}=\left( \frac{(a^{i})^{2}}{2}+A^{0}\right) _{x},
\end{equation*}%
if all field variables $b^{k}$ vanish, and $A^{0}$ becomes a function of all
rest field variables $a^{n}$ only. Let us replace $a^{i}$ by $q(x,t,\lambda
) $, where $\lambda $ is a parameter. It means $a^{i}=q(x,t,\xi ^{i})$,
where $\xi ^{i}$ are arbitrary constants. Then (\ref{2})%
\begin{equation*}
q_{t}=qq_{x}+A_{x}^{0}
\end{equation*}%
by a semi-hodograph transformation $q(x,t,\lambda )\rightarrow \lambda
(x,t,q)$ reduces to the linear equation%
\begin{equation*}
\lambda _{t}=q\lambda _{x}-\lambda _{q}A_{x}^{0},
\end{equation*}%
which is known as the Vlasov kinetic equation (see \cite{zakh}; or the
collisionless Boltzmann equation, see \cite{Gibbons}). Suppose $\lambda
(x,t,q)$ is a function $\lambda (q,A^{0},A^{1},...)$, where all moments $%
A^{k}(x,t)$ satisfy Benney hydrodynamic chain (\ref{1}). Since we suppose
all moments $A^{k}$ are independent, one can obtain an infinite series of
equations%
\begin{equation}
\partial _{k}\lambda =q^{-k}\partial _{0}\lambda ,\text{ }k=0,1,2,...,
\label{sol}
\end{equation}%
where $\partial _{k}\equiv \partial /\partial A^{k}$, and ($\partial
_{q}\equiv \partial /\partial q$)%
\begin{equation}
\partial _{0}\lambda =\left( q-\sum_{m=0}^{\infty }\frac{mA^{m-1}}{q^{m}}%
\right) ^{-1}\partial _{q}\lambda .  \label{ha}
\end{equation}%
A solution of (\ref{sol}) is given by\footnote{%
It is well known that a general solution of the above linear equation is
parameterized by one arbitrary function of a single variable $\tilde{\lambda}%
(\lambda )$. However, in this approach, an existence of \textit{any}
solution is essential.}%
\begin{equation}
\lambda =B_{1}(q)\sum_{m=0}^{\infty }\frac{A^{m}}{q^{m+1}}+B_{2}(q),
\label{ho}
\end{equation}%
where $B_{1}(q)$ and $B_{2}(q)$ are not determined yet functions. However, a
substitution (\ref{ho}) into (\ref{ha}) yields $B_{1}(q)=1$ and $B_{2}(q)=q$%
. Then (\ref{ho}) becomes nothing else but an inverse series to (\ref{3}).
Thus, we conclude that a symptom of an integrability of hydrodynamic chains
is an existence of a Riemann mapping $\lambda (q,A^{0},A^{1},...)$
connecting with Vlasov type kinetic equation (see below). In this paper, we
utilize this property for a classification of integrable hydrodynamic chains.

This moment decomposition approach can be extended on a wide class of
hydrodynamic chains (cf. (\ref{lin}))%
\begin{equation}
A_{t}^{k}=\sum_{n=0}^{K}f_{n}A_{x}^{k+n}+\sum_{m=0}^{M}\left(
\sum_{n=0}^{K}A^{k+n}s_{nm}+k\sum_{n=-1}^{K}A^{k+n}w_{nm}\right) A_{x}^{m}
\label{lina}
\end{equation}%
where $K$ and $M$ are arbitrary natural numbers, all functions $%
f_{i},s_{jk},w_{lp}$ depend on first $M+1$ moments (if $%
K=1,M=2,s_{0,2}=0,s_{1,2}=0$ and $%
f_{0},f_{1},s_{0,0},s_{0,1},s_{1,0},s_{1,1} $ depend just on two first
moments $A^{0},A^{1}$, these hydrodynamic chains reduce to (\ref{lin})).
Indeed, (\ref{lina}) reduces to $N$ separate expressions for each index $i$
(let remind that $N$ is arbitrary)%
\begin{eqnarray*}
(a^{i})^{k}b_{t}^{i}+k(a^{i})^{k-1}b^{i}a_{t}^{i}
&=&%
\sum_{n=0}^{K}f_{n}[(a^{i})^{k+n}b_{x}^{i}+(k+n)(a^{i})^{k+n-1}b^{i}a_{x}^{i}]
\\
&&+\sum_{m=0}^{M}\left(
\sum_{n=0}^{K}(a^{i})^{k+n}b^{i}s_{nm}+k%
\sum_{n=-1}^{K}(a^{i})^{k+n}b^{i}w_{nm}\right) A_{x}^{m},
\end{eqnarray*}%
due to a substitution (\ref{zakh}) in \textit{moments equipped by the index} 
$k$ \textit{only}. Moreover, the above $N$ expressions (due to their linear
explicit dependence on a discrete variable $k$) can be split on two parts%
\begin{equation*}
b_{t}^{i}=%
\sum_{n=0}^{K}f_{n}[(a^{i})^{n}b_{x}^{i}+n(a^{i})^{n-1}b^{i}a_{x}^{i}]+%
\sum_{m=0}^{M}\sum_{n=0}^{K}(a^{i})^{n}b^{i}s_{nm}A_{x}^{m},
\end{equation*}%
\begin{equation}
a_{t}^{i}=\sum_{n=0}^{K}f_{n}\cdot
(a^{i})^{n}a_{x}^{i}+\sum_{m=0}^{M}%
\sum_{n=-1}^{K}(a^{i})^{n+1}w_{nm}A_{x}^{m}.  \label{reduk}
\end{equation}%
As in the previous case, this $2N$ component hydrodynamic type system
possesses $N$ component reduction (\ref{reduk}), where all moments $A^{m}$
and all variable coefficients $w_{ik}$ depend on $N$ field variables $a^{n}$
only.

Our main observation successfully utilized in this approach is that \textbf{%
\textit{integrable} hydrodynamic chain} (\ref{lina}) \textbf{is associated
with the auxiliary equation}%
\begin{equation}
q_{t}=\sum_{n=0}^{K}f_{n}q^{n}q_{x}+\sum_{m=0}^{M}%
\sum_{n=-1}^{K}q^{n+1}w_{nm}A_{x}^{m},  \label{extra}
\end{equation}%
which obtains due to a formal replacement $a^{i}\rightarrow q$ in (\ref%
{reduk}). It means, that equation (\ref{extra}) is compatible with
hydrodynamic chain (\ref{lina}), where function $q$ must depend on moments $%
A^{k}(x,t)$ and the parameter $\lambda $. The semi-hodograph transformation $%
q(x,t,\lambda )\leftrightarrow \lambda (x,t,q)$ reduces (\ref{extra}) to the
linear equation%
\begin{equation}
\lambda _{t}-\sum_{n=0}^{K}f_{n}q^{n}\lambda
_{x}+\sum_{m=0}^{M}\sum_{n=-1}^{K}q^{n+1}w_{nm}A_{x}^{m}\lambda _{q}=0,
\label{linn}
\end{equation}%
which we call the Vlasov type kinetic equation (cf. \cite{OPS}). The
function $\lambda (x,t,q)$ depends on $x,t$ implicitly via an explicit
dependence on moments $A^{k}(x,t)$. We shall call hydrodynamic chain (\ref%
{lina}) \textit{integrable} if a Riemann mapping $\lambda
(q,A^{0},A^{1},...) $ connecting Vlasov type kinetic equation (\ref{linn})
with (\ref{lina}) exists.

\textbf{Examples}: Hamiltonian hydrodynamic chains associated with the
Kupershmidt--Manin Poisson bracket (see \cite{GibRai}; $h_{n}\equiv \partial
h/\partial A^{n},h_{nm}\equiv \partial ^{2}h/\partial A^{n}\partial A^{m}$)%
\begin{equation*}
A_{t}^{k}=\sum_{n=0}^{M-1}(n+1)h_{n+1}A_{x}^{k+n}+\sum_{m=0}^{M}\left(
\sum_{n=0}^{M-1}(n+1)A^{k+n}h_{n+1,m}+k\sum_{n=-1}^{M-1}A^{k+n}h_{n+1,m}%
\right) A_{x}^{m}
\end{equation*}%
are connected with the Vlasov type kinetic equation%
\begin{equation*}
\lambda _{t}-\sum_{n=0}^{M-1}(n+1)h_{n+1}q^{n}\lambda
_{x}+\sum_{m=0}^{M}\sum_{n=-1}^{M-1}q^{n+1}h_{n+1,m}A_{x}^{m}\lambda _{q}=0,
\end{equation*}%
where the Hamiltonian is given by $\mathbf{H}=\int
h(A^{0},A^{1},...,A^{M})dx $ and (see (\ref{lina}))%
\begin{equation*}
f_{n}=(n+1)h_{n+1},\text{\ }K=M-1,\text{\ }s_{nm}=(n+1)h_{n+1,m},\text{\ }%
w_{nm}=h_{n+1,m}.
\end{equation*}%
Hamiltonian hydrodynamic chains associated with the Kupershmidt Poisson
brackets (see \cite{FKMP} and \cite{KuperNorm})%
\begin{equation*}
A_{t}^{k}=\sum_{n=0}^{M}(\alpha n+\beta
)h_{n}A_{x}^{k+n}+\sum_{m=0}^{M}\left( \sum_{n=0}^{M}(\alpha n+2\beta
)A^{k+n}h_{nm}+\alpha k\sum_{n=0}^{M}A^{k+n}h_{nm}\right) A_{x}^{m}
\end{equation*}%
are connected with the Vlasov type kinetic equation%
\begin{equation*}
\lambda _{t}-\sum_{n=0}^{M}(\alpha n+\beta )h_{n}q^{n}\lambda _{x}+\alpha
\sum_{m=0}^{M}\sum_{n=0}^{M}q^{n+1}h_{nm}A_{x}^{m}\lambda _{q}=0
\end{equation*}%
where the Hamiltonian is given by $\mathbf{H}=\int
h(A^{0},A^{1},...,A^{M})dx $ and (see (\ref{lina}))%
\begin{equation*}
f_{n}=(\alpha n+\beta )h_{n},\text{\ }K=M,\text{\ }s_{nm}=(\alpha n+2\beta
)h_{nm},\text{\ }w_{nm}=\alpha h_{nm},\text{ \ }w_{-1,m}=0.
\end{equation*}

Without loss of generality and for simplicity let us consider a hydrodynamic
chain written in the form%
\begin{equation}
A_{t}^{k}=\sum_{n=0}^{1}f_{n}A_{x}^{k+n}+\sum_{m=0}^{M}\left(
\sum_{n=0}^{1}A^{k+n}s_{nm}+k\sum_{n=-1}^{1}A^{k+n}w_{nm}\right) A_{x}^{m}.
\label{link}
\end{equation}%
If this hydrodynamic chain is integrable, then also all its higher commuting
flows belong to the general class determined by (\ref{lina}) with
appropriate choices natural numbers $K$ and $M$.

\textbf{Lemma}: \textit{The coefficient} $f_{1}$ \textit{can be fixed to the
unity by the invertible point transformation }$\tilde{A}%
^{k}=(f_{1})^{k}A^{k} $\textit{, then the coefficient} $f_{0}$ \textit{can
be eliminated by the invertible point transformation}\footnote{%
Similar transformations preserving the Kupershmidt--Manin Poisson bracket
were considered in \cite{FerMarsh}, but with \textit{constant} coefficients $%
f_{0}$ and $f_{1}$.}%
\begin{equation*}
\tilde{A}^{k}=\sum_{m=0}^{k}\binom{k}{m}(f_{0})^{k-m}A^{m},
\end{equation*}%
\textit{where} $\binom{k}{m}$ \textit{is a binomial coefficient}. \textit{If}
$\partial _{M}f^{1}=0$ \textit{and} $\partial _{M}f^{0}=0$, \textit{then
hydrodynamic chain} (\ref{link}) \textit{reduces to the \textbf{canonical}
form}%
\begin{equation}
A_{t}^{k}=A_{x}^{k+1}+\sum_{m=0}^{M}\left(
\sum_{n=0}^{1}A^{k+n}s_{nm}+k\sum_{n=-1}^{1}A^{k+n}w_{nm}\right) A_{x}^{m};
\label{linp}
\end{equation}%
\textit{if} $\partial _{M}f^{1}\neq 0$ \textit{or} $\partial _{M}f^{0}\neq 0$%
, \textit{then} (\ref{link}) \textit{reduces to} (\ref{linp}), \textit{but} $%
M$ \textit{replaces by} $M+1$, \textit{correspondingly}.

\textbf{Proof}: Hydrodynamic chain (\ref{link}) is associated with the
reduced version of (\ref{extra})%
\begin{equation*}
q_{t}=(f_{1}q+f_{0})q_{x}+\sum_{m=0}^{M}%
\sum_{n=-1}^{1}q^{n+1}w_{nm}A_{x}^{m}.
\end{equation*}%
Thus, the transformation $\tilde{q}=f_{1}q+f_{0}$ reduces the above equation
to a more simple case with $f_{1}=1$ and $f_{0}=0$. Corresponding Zakharov's
moment decomposition (\ref{zakh}) transforms accordingly%
\begin{equation}
\tilde{A}^{k}=\sum_{m=0}^{N}(f_{1}a^{m}+f_{0})^{k}b^{m}.  \label{mom}
\end{equation}%
This is nothing else but a linear combination of aforementioned
transformations. This point transformation $\tilde{A}^{0}=A^{0},\tilde{A}%
^{1}=f_{1}A^{0}+(f_{0})^{2},\tilde{A}%
^{2}=(f_{1})^{2}A^{2}+2f_{0}f_{1}A^{1}+(f_{0})^{3},$ $...,\tilde{A}%
^{M}=(f_{1})^{M}A^{M}+Mf_{0}(f_{1})^{M-1}A^{M-1}+...+(f_{0})^{M+1},...$
cannot be inverted to a similar form due to complexity of functions $%
f_{0}(A^{0},A^{1},...,A^{M})$ and $f_{1}(A^{0},A^{1},...,A^{M})$. Just
higher moments $A^{M+k}(\tilde{A}^{0},\tilde{A}^{1},...,\tilde{A}^{M+k})$
became \textit{linear} expressions with respect to higher moments $\tilde{A}%
^{M+1},\tilde{A}^{M+2},...$

On the other hand, $2N$ component hydrodynamic type system (\ref{reduk})%
\begin{equation*}
b_{t}^{i}=(f_{1}a^{i}+f_{0})b_{x}^{i}+f_{1}b^{i}a_{x}^{i}+b^{i}%
\sum_{m=0}^{M}\sum_{n=0}^{1}(a^{i})^{n}s_{nm}A_{x}^{m},\text{ }%
a_{t}^{i}=(f_{1}a^{i}+f_{0})a_{x}^{i}+\sum_{m=0}^{M}%
\sum_{n=-1}^{1}(a^{i})^{n+1}w_{nm}A_{x}^{m}
\end{equation*}%
under the aforementioned transformation $c^{i}=f_{1}a^{i}+f_{0}$ reduces to%
\begin{equation}
b_{t}^{i}=c^{i}b_{x}^{i}+b^{i}c_{x}^{i}+b^{i}\sum_{m=0}^{M}%
\sum_{n=0}^{1}(c^{i})^{n}\bar{s}_{nm}A_{x}^{m},\text{ \ }%
c_{t}^{i}=c^{i}c_{x}^{i}+\sum_{m=0}^{M+1}\sum_{n=-1}^{1}(c^{i})^{n+1}\bar{w}%
_{nm}A_{x}^{m},  \label{5}
\end{equation}%
where%
\begin{equation*}
\bar{s}_{1m}=\frac{s_{1m}}{f_{1}}-\partial _{m}\ln f_{1},\text{ \ }\bar{s}%
_{0m}=s_{0m}-\frac{f_{0}}{f_{1}}s_{1m}+f_{0}\partial _{m}\ln f_{1}-\partial
_{m}f_{0},\text{ \ }
\end{equation*}%
\begin{equation*}
\bar{w}_{1m}=\frac{w_{1m}}{f_{1}}-\partial _{m}\ln f_{1},\text{ \ }\bar{w}%
_{0,M+1}=\partial _{M}f_{1},\text{ \ }\bar{w}_{-1,M+1}=f_{1}\partial
_{M}f_{0}-f_{0}\partial _{M}f_{1},
\end{equation*}%
\begin{equation*}
\bar{w}_{0m}=w_{0m}+(1-\delta _{m,0})\partial _{m-1}f_{1}-\partial
_{m}f_{0}+2f_{0}\partial _{m}\ln f_{1}-\frac{2f_{0}w_{1m}}{f_{1}}
\end{equation*}%
\begin{equation*}
+\sum_{p=0}^{M}\left(
\sum_{n=0}^{1}A^{p+n}s_{nm}+\sum_{n=-1}^{1}pA^{p+n}w_{nm}\right) \partial
_{p}\ln f_{1},
\end{equation*}%
\begin{equation*}
\bar{w}_{-1m}=(1-\delta _{m,0})(f_{1}\partial _{m-1}f_{0}-f_{0}\partial
_{m-1}f_{1})+f_{0}\partial _{m}f_{0}-(f_{0})^{2}\partial _{m}\ln f_{1}+\frac{%
(f_{0})^{2}w_{1m}}{f_{1}}-f_{0}w_{0m}+f_{1}w_{-1m}
\end{equation*}%
\begin{equation*}
-f_{0}\sum_{p=0}^{M}\left(
\sum_{n=0}^{1}A^{p+n}s_{nm}+\sum_{n=-1}^{1}pA^{p+n}w_{nm}\right) \partial
_{p}\ln f_{1}+\sum_{p=0}^{M}\left(
\sum_{n=0}^{1}A^{p+n}s_{nm}+\sum_{n=-1}^{1}pA^{p+n}w_{nm}\right) \partial
_{p}f_{0}.
\end{equation*}%
Due to (\ref{mom}), (\ref{5}) can be written in the final form%
\begin{equation*}
b_{t}^{i}=c^{i}b_{x}^{i}+b^{i}c_{x}^{i}+b^{i}\sum_{m=0}^{M}%
\sum_{n=0}^{1}(c^{i})^{n}\tilde{s}_{nm}\tilde{A}_{x}^{m},\text{ \ }%
c_{t}^{i}=c^{i}c_{x}^{i}+\sum_{m=0}^{M+1}\sum_{n=-1}^{1}(c^{i})^{n+1}\tilde{w%
}_{nm}\tilde{A}_{x}^{m},
\end{equation*}%
where coefficients $\tilde{s}_{ij}$ and $\tilde{w}_{kl}$ are expressed via
new moments $\tilde{A}^{n}$. This is nothing else but a hydrodynamic
reduction of hydrodynamic chain (\ref{linp})%
\begin{equation*}
\tilde{A}_{t}^{k}=\tilde{A}_{x}^{k+1}+\sum_{m=0}^{M+1}\left( \sum_{n=0}^{1}%
\tilde{A}^{k+n}\tilde{s}_{nm}+k\sum_{n=-1}^{1}\tilde{A}^{k+n}\tilde{w}%
_{nm}\right) \tilde{A}_{x}^{m}.
\end{equation*}%
Thus, Lemma is proved.

\textbf{Example}: The remarkable Kupershmidt hydrodynamic chain (see \cite%
{KuperNorm}, \cite{MaksKuper})%
\begin{equation*}
A_{t}^{k}=A_{x}^{k+1}+\beta A^{0}A_{x}^{k}+(k+\gamma )A^{k}A_{x}^{0},\text{ }%
k=0,1,...
\end{equation*}%
reduces to canonical form (\ref{linp})%
\begin{equation*}
\tilde{A}_{t}^{k}=\tilde{A}_{x}^{k+1}+[(1-\beta )k+\gamma -\beta ]\tilde{A}%
^{k}\tilde{A}_{x}^{0}+\beta k\tilde{A}^{k-1}\left( \tilde{A}^{1}+\frac{%
\gamma -\beta -1}{2}(\tilde{A}^{0})^{2}\right) _{x},\text{ }k=0,1,...
\end{equation*}

Thus, we can investigate an integrability of hydrodynamic chain (\ref{link})
written in a more convenient form (\ref{linp}) instead (\ref{link}). In such
a case, (\ref{extra}) reduces to%
\begin{equation}
q_{t}=qq_{x}+\sum_{m=0}^{M}\sum_{n=-1}^{1}q^{n+1}w_{nm}A_{x}^{m}.
\label{ext}
\end{equation}%
A consistency of (\ref{ext}) with (\ref{linp}) leads to an infinite set of
equations%
\begin{equation}
\partial _{M+k}q=q^{-k}\partial _{M}q,\text{ \ }k=0,1,2,...  \label{simple}
\end{equation}%
and $M+1$ equations ($m=0,1,2,...,M$) reduces by virtue of (\ref{simple}) to%
\begin{equation*}
(1-\delta _{m,0})\partial _{m-1}q+\sum_{k=0}^{M-1}\left(
\sum_{n=0}^{1}s_{nm}A^{k+n}+k\sum_{n=-1}^{1}w_{nm}A^{k+n}\right) \partial
_{k}q+\Sigma _{m}\cdot \partial _{M}q=\sum_{n=-1}^{1}w_{nm}q^{n+1}+q\partial
_{m}q,
\end{equation*}%
where $M+1$ infinite sums are determined by%
\begin{equation}
\Sigma _{m}=\sum_{k=0}^{\infty }\left(
\sum_{n=0}^{1}s_{nm}A^{M+n+k}+(M+k)\sum_{n=-1}^{1}w_{nm}A^{M+n+k}\right) 
\frac{1}{q^{k}}.  \label{sigma}
\end{equation}%
However, all these infinite sums can be reduced to a sole sum only (see
below). A consistency of the Vlasov type kinetic equation (cf. (\ref{linn}),
see also (\ref{ext}))%
\begin{equation*}
\lambda _{t}=q\lambda
_{x}-\sum_{m=0}^{M}\sum_{n=-1}^{1}q^{n+1}w_{nm}A_{x}^{m}\lambda _{q}
\end{equation*}%
with (\ref{linp}) yields an infinite set of equations (which is equivalent
to (\ref{simple}) due to the transformation $\partial _{k}q=-\partial
_{k}\lambda /\partial _{q}\lambda $) 
\begin{equation*}
\partial _{M+k}\lambda =q^{-k}\partial _{M}\lambda ,\text{ \ }k=0,1,2,...,
\end{equation*}%
whose solution is given by%
\begin{equation*}
\lambda =B_{1}(q,A^{0},A^{1},...,A^{M-1})[\Sigma
+B_{2}(q,A^{0},A^{1},...,A^{M-1})],
\end{equation*}%
where%
\begin{equation}
\Sigma =\sum_{p=0}^{\infty }\frac{A^{p}}{q^{p+1}},  \label{sig}
\end{equation}%
while other $M+1$ equations%
\begin{equation*}
(1-\delta _{m,0})\partial _{m-1}\lambda +\sum_{k=0}^{M-1}\left(
\sum_{n=0}^{1}s_{nm}A^{k+n}+k\sum_{n=-1}^{1}w_{nm}A^{k+n}\right) \partial
_{k}\lambda +\Sigma _{m}\cdot \partial _{M}\lambda
+\sum_{n=-1}^{1}w_{nm}q^{n+1}\partial _{q}\lambda =q\partial _{m}\lambda ,
\end{equation*}%
reduce to a linear system\footnote{%
This linear system does not contain a part proportional to $\partial
_{q}\Sigma $, because its corresponding coefficient vanishes authomatically.}
$G_{m}(q,A^{0},A^{1},...,A^{M})\Sigma +Q_{m}(q,A^{0},A^{1},...,A^{M})=0$ due
to (\ref{sigma}) expresses via (\ref{sig})%
\begin{eqnarray*}
\Sigma _{m} &=&\left(
\sum_{n=0}^{1}s_{nm}q^{M+n+1}-\sum_{n=-1}^{1}(n+1)w_{nm}q^{M+n+1}\right)
\Sigma -\sum_{n=-1}^{1}w_{nm}q^{M+n+2}\partial _{q}\Sigma \\
&& \\
&&+\sum_{n=-1}^{1}w_{nm}\sum_{p=0}^{M+n-1}\frac{n-p}{q^{p-n-M}}%
A^{p}-\sum_{n=0}^{1}s_{nm}\sum_{p=0}^{M+n-1}\frac{A^{p}}{q^{p-n-M}}.
\end{eqnarray*}%
Since both coefficients $G_{m}$ and $Q_{m}$ must vanish independently, a
full $2M+2$ component system can be split on the two $M+1$ component
sub-systems of linear equations%
\begin{equation*}
q\partial _{m}\ln B_{1}+(\delta _{m,0}-1)\partial _{m-1}\ln
B_{1}-\sum_{k=0}^{M-1}\left(
\sum_{n=0}^{1}s_{nm}A^{k+n}+k\sum_{n=-1}^{1}w_{nm}A^{k+n}\right) \partial
_{k}\ln B_{1}
\end{equation*}%
\begin{equation}
-\sum_{n=-1}^{1}w_{nm}q^{n+1}\partial _{q}\ln
B_{1}=\sum_{n=0}^{1}[s_{nm}-(n+1)w_{nm}]q^{n}  \label{uno}
\end{equation}%
\qquad 
\begin{equation*}
\text{and}
\end{equation*}%
\begin{equation*}
q\partial _{m}B_{2}+(\delta _{m,0}-1)\partial
_{m-1}B_{2}-\sum_{k=0}^{M-1}\left(
\sum_{n=0}^{1}s_{nm}A^{k+n}+k\sum_{n=-1}^{1}w_{nm}A^{k+n}\right) \partial
_{k}B_{2}-\sum_{n=-1}^{1}w_{nm}q^{n+1}\partial _{q}B_{2}
\end{equation*}%
\begin{equation*}
+\sum_{n=0}^{1}[s_{nm}-(n+1)w_{nm}]q^{n}B_{2}=A^{0}(s_{1m}-w_{1m})-\delta
_{m,0}.
\end{equation*}

All derivatives of functions $B_{1}$ and $B_{2}$ can be expressed from this
linear system with variable coefficients $s_{ij}$ and $w_{kl}$. A
consistency of these derivatives leads to an overdetermined system in
partial derivatives on $s_{ij}$ and $w_{kl}$ with respect to moments $%
A^{0},A^{1},...,A^{M-1}$ and $q$, while a dependence on the highest moment $%
A^{M}$ can be found by a straightforward differentiation of linear system (%
\ref{uno}) written in the matrix\footnote{%
a determinant of this $(M+1)\times (M+1)$ matrix is a polynomial of degree $%
M+2$ with respect to $q$, except some special cases, like $w_{1,M}=0$, which
should be considered separately.} form%
\begin{equation*}
\left( 
\begin{array}{lllll}
q+\ast & \ast & ... & \ast & -w_{1,1}q^{2}-w_{0,1}q+\ast \\ 
\ast & q+\ast & ... & \ast & -w_{1,2}q^{2}-w_{0,2}q+\ast \\ 
... & ... & ... & ... & ... \\ 
\ast & \ast & ... & q+\ast & -w_{1,M-1}q^{2}-w_{0,M-1}q+\ast \\ 
\ast & \ast & ... & \ast & -w_{1,M}q^{2}-w_{0,M}q+\ast%
\end{array}%
\right) \left( 
\begin{array}{l}
\partial _{0}\ln B^{1} \\ 
\partial _{1}\ln B^{1} \\ 
... \\ 
\partial _{M-1}\ln B^{1} \\ 
\partial _{q}\ln B^{1}%
\end{array}%
\right) =\left( 
\begin{array}{l}
\tilde{w}_{0}q+\ast \\ 
\tilde{w}_{1}q+\ast \\ 
... \\ 
\tilde{w}_{M-1}q+\ast \\ 
\tilde{w}_{M}q+\ast%
\end{array}%
\right) ,
\end{equation*}%
where $\tilde{w}_{k}=s_{1k}-2w_{1k}$ and the mark \textquotedblleft $\ast $%
\textquotedblright\ means elements independent on $q$. Indeed, such a
differential consequence is given by%
\begin{equation*}
\left( 
\begin{array}{lllll}
\ast & \ast & ... & \ast & -w_{1,1}^{\prime }q^{2}-w_{0,1}^{\prime }q+\ast
\\ 
\ast & \ast & ... & \ast & -w_{1,2}^{\prime }q^{2}-w_{0,2}^{\prime }q+\ast
\\ 
... & ... & ... & ... & ... \\ 
\ast & \ast & ... & \ast & -w_{1,M-1}^{\prime }q^{2}-w_{0,M-1}^{\prime
}q+\ast \\ 
\ast & \ast & ... & \ast & -w_{1,M}^{\prime }q^{2}-w_{0,M}^{\prime }q+\ast%
\end{array}%
\right) \left( 
\begin{array}{l}
\partial _{0}\ln B^{1} \\ 
\partial _{1}\ln B^{1} \\ 
... \\ 
\partial _{M-1}\ln B^{1} \\ 
\partial _{q}\ln B^{1}%
\end{array}%
\right) =\left( 
\begin{array}{l}
\tilde{w}_{0}^{\prime }q+\ast \\ 
\tilde{w}_{1}^{\prime }q+\ast \\ 
... \\ 
\tilde{w}_{M-1}^{\prime }q+\ast \\ 
\tilde{w}_{M}^{\prime }q+\ast%
\end{array}%
\right) ,
\end{equation*}%
where the mark \textquotedblleft $^{\prime }$\textquotedblright\ means a
partial derivative with respect to the moment $A^{M}$.

\textbf{Lemma}: \textit{Any row of the above linear system}%
\begin{equation}
\left( 
\begin{array}{llllll}
\ast , & \ast , & ..., & \ast , & -w_{1,m}^{\prime }q^{2}-w_{0,m}^{\prime
}q+\ast , & \tilde{w}_{m}^{\prime }q+\ast%
\end{array}%
\right)  \label{lane}
\end{equation}%
\textit{is proportional to the last row from the previous linear system}%
\begin{equation}
\left( 
\begin{array}{llllll}
\ast , & \ast , & ..., & \ast , & -w_{1,M}q^{2}-w_{0,M}q+\ast , & \tilde{w}%
_{M}q+\ast%
\end{array}%
\right) .  \label{lan}
\end{equation}

\textbf{Proof}: Indeed, let us consider the linear system%
\begin{equation*}
\left( 
\begin{array}{llllll}
q+\ast & \ast & ... & \ast & -w_{1,1}q^{2}-w_{0,1}q+\ast & -\tilde{w}%
_{0}q-\ast \\ 
\ast & q+\ast & ... & \ast & -w_{1,2}q^{2}-w_{0,2}q+\ast & -\tilde{w}%
_{1}q-\ast \\ 
... & ... & ... & ... & ... & ... \\ 
\ast & \ast & ... & q+\ast & -w_{1,M-1}q^{2}-w_{0,M-1}q+\ast & -\tilde{w}%
_{M-1}q-\ast \\ 
\ast & \ast & ... & \ast & -w_{1,M}q^{2}-w_{0,M}q+\ast & -\tilde{w}_{M}q-\ast
\\ 
\ast & \ast & ... & \ast & -w_{1,m}^{\prime }q^{2}-w_{0,m}^{\prime }q+\ast & 
-\tilde{w}_{m}^{\prime }q-\ast%
\end{array}%
\right) \left( 
\begin{array}{l}
\partial _{0}B^{1} \\ 
\partial _{1}B^{1} \\ 
... \\ 
\partial _{M-1}B^{1} \\ 
\partial _{q}B^{1} \\ 
B^{1}%
\end{array}%
\right) =0,
\end{equation*}%
determined by the $(M+2)\times (M+2)$ matrix incorporating all rows of the
original linear system and any row from its differential consequence. A
determinant of this matrix equals zero for nontrivial solutions $B^{1}$.
Thus, the last row (see (\ref{lane})) must be a linear combination of all
other rows. However, most of them ($m=0,1,...,M-1$) contain an element $%
q+\ast $, which does not exist in first $M$ entries of this last row. Thus,
the last row cannot be expressed via these higher flows except the row with
the number $M$ (see (\ref{lan})). It means, that all elements of these 
\textit{two} rows must be proportional to each other. Lemma is proved.

Thus, the full set of equations is given by ($n=0,1,...,M-2$)%
\begin{equation}
\frac{\beta _{M}^{\prime }}{\beta _{M}}=\frac{\delta _{M}^{\prime }}{\delta
_{M}}=\frac{(\epsilon _{M}^{M-1})^{\prime }}{1+\epsilon _{M}^{M-1}}=\frac{%
(\epsilon _{M}^{n})^{\prime }}{\epsilon _{M}^{n}},  \label{odin}
\end{equation}%
\begin{equation}
\frac{\beta _{m}^{\prime }}{\beta _{M}}=\frac{\delta _{m}^{\prime }}{\delta
_{M}}=\frac{(\epsilon _{m}^{M-1})^{\prime }}{1+\epsilon _{M}^{M-1}}=\frac{%
(\epsilon _{m}^{n})^{\prime }}{\epsilon _{M}^{n}},\text{ }m=0,1,...,M-1,
\label{tre}
\end{equation}%
where 
\begin{eqnarray*}
\beta _{m} &=&s_{0,m}-w_{0,m}+(s_{1,m}-2w_{1,m})q, \\
&& \\
\epsilon _{m}^{n}
&=&nw_{-1,m}A^{n-1}+(s_{0,m}+nw_{0,m})A^{n}+(s_{1,m}+nw_{1,m})A^{n+1}, \\
&& \\
\delta _{m} &=&w_{-1,m}+w_{0,m}q+w_{1,m}q^{2}.
\end{eqnarray*}%
All these equations can be subsequently integrated. Indeed, the first ratio
in (\ref{odin})%
\begin{equation*}
\frac{\beta _{M}^{\prime }}{\beta _{M}}=\frac{\delta _{M}^{\prime }}{\delta
_{M}}
\end{equation*}%
is nothing else but a cubic polynomial with respect to $q$. Since $q$ is
arbitrary, all four coefficients must vanish independently. A general
solution of corresponding four ordinary differential equations (with respect
to $A^{M}$ only) is given by%
\begin{equation*}
s_{0,M}=\left( r_{0}-M+1\right) w_{1,M},\text{ \ \ }s_{1,M}=\left(
r_{1}-M+1\right) w_{1,M},
\end{equation*}%
\begin{equation}
w_{0,M}=u_{0}w_{1,M},\text{ \ \ }w_{-1,M}=u_{-1}w_{1,M},  \label{he}
\end{equation}%
where functions $r_{0},u_{0},r_{1},u_{-1}$ depend on first $M$ moments $%
A^{0},A^{1},...,A^{M-1}$. An integration of the second ratio in (\ref{odin})%
\begin{equation*}
\frac{\delta _{M}^{\prime }}{\delta _{M}}=\frac{(\epsilon
_{M}^{M-1})^{\prime }}{1+\epsilon _{M}^{M-1}}
\end{equation*}%
leads to%
\begin{equation}
w_{1,M}=-\frac{1}{\sigma +A_{M}r_{1}},  \label{hi}
\end{equation}%
where the function $\sigma $ depends on first $M$ moments $A^{0},A^{1}\dots
,A^{M-1}$. An integration of the first ratio in (\ref{tre})%
\begin{equation*}
\frac{\beta _{m}^{\prime }}{\beta _{M}}=\frac{\delta _{m}^{\prime }}{\delta
_{M}}
\end{equation*}%
leads to%
\begin{equation}
w_{-1,m}=u_{-1}w_{1,m}+\gamma _{-1,m},\qquad w_{0,m}=u_{0}w_{1,m}+\gamma
_{0,m},  \label{hy}
\end{equation}%
\begin{equation*}
s_{1,m}=\left( r_{1}-M+1\right) w_{1,m}+\rho _{1,m},\qquad s_{0,m}=\left(
r_{0}-M+1\right) w_{1,m}+\rho _{0,m},
\end{equation*}%
where functions $\gamma _{-1,m},\gamma _{0,m},\rho _{0,m},\rho _{1,m}$
depend on first $M$ moments $A^{0},A^{1},...,A^{M-1}$. An integration of the
second ratio in (\ref{tre})%
\begin{equation*}
\frac{\delta _{m}^{\prime }}{\delta _{M}}=\frac{(\epsilon
_{m}^{M-1})^{\prime }}{1+\epsilon _{M}^{M-1}}
\end{equation*}%
leads to%
\begin{equation}
w_{1,m}=\frac{\omega _{m}-A_{M}\rho _{1,m}}{\sigma +A_{M}\text{ }r_{1}},
\label{hu}
\end{equation}%
where functions $\omega _{m}$ depend on first $M$ moments $%
A^{0},A^{1},...,A^{M-1}$. It is easy to see, that all other ratios in (\ref%
{odin}) and (\ref{tre}) are fulfilled by virtue of (\ref{he}), (\ref{hi}), (%
\ref{hy}), (\ref{hu}).

In the next Section, a more deep analysis is presented for hydrodynamic
chain (\ref{lin}) written in form (\ref{linp}).

\section{Canonical variables}

The function $B_{1}$ depends on first $M$ moments $A^{0},A^{1},A^{M-1}$
only, but coefficients of linear system (\ref{uno}) depend also on $A^{M}$
explicitly via (\ref{he}), (\ref{hi}), (\ref{hy}), (\ref{hu}). Thus, each
derivative of $\ln B_{1}$ can be expressed as a ratio of two polynomials
with respect to $q$. In a general case (if $w_{1,M}\neq 0$), the common
denominator is a polynomial of a degree $M+2$. All numerators are
polynomials of the same degree, except a numerator of derivative $\ln B_{1}$
with respect to $q$. Its degree is $M+1$. Let us introduce roots $%
q_{k}(A^{0},A^{1},...,A^{M-1})$ of this polynomial as \textbf{basic field
variables} for further computations. In such a case,%
\begin{equation}
\partial _{q}\ln B_{1}=-\sum_{m=1}^{M+2}\frac{\alpha _{m}}{q-q_{m}},
\label{bi}
\end{equation}%
where $\alpha _{m}(A^{0},A^{1},...,A^{M-1})$ are not yet determined
functions. It means, that%
\begin{equation}
B_{1}=\alpha _{0}\overset{M+2}{\underset{m=1}{\prod }}(q-q_{m})^{-\alpha
_{m}},  \label{ba}
\end{equation}%
where $\alpha _{0}(A^{0},A^{1},...,A^{M-1})$ is not yet determined function.
A substitution (\ref{ba}) back to the first derivative of $\ln B_{1}$ with
respect to $q$ allows to express few (not all) variable coefficients ($%
r_{0},u_{0},r_{1},u_{-1},\sigma ,$ $\gamma _{-1,m},\gamma _{0,m},\rho
_{0,m},\rho _{1,m},\omega _{m}$, see the previous Section) via new field
variables $q_{k}$. Moreover, $\alpha _{0}$ and all other $\alpha _{m}$ must
be constant parameters (this is a consequence of an absence of logarithmic
terms in derivatives of $\ln B_{1}$ with respect to moments $A^{k}$); $r_{1}$
is constant due to the constraint 
\begin{equation}
\sum_{m=1}^{M+2}\alpha _{m}=\frac{1}{r_{1}}-M-1,  \label{er}
\end{equation}%
following from comparison of r.h.s. in (\ref{bi}) with a corresponding
expression from linear system (\ref{uno}). Without loss of generality, one
can fix $\alpha _{0}$ on the unity. The compatibility conditions $\partial
_{k}(\partial _{q}\ln B_{1})=\partial _{q}(\partial _{k}\ln B_{1}),\partial
_{k}(\partial _{n}\ln B_{1})=\partial _{n}(\partial _{k}\ln B_{1})$ imply to
explicit relationships between some coefficients as well as dependencies $%
\partial _{k}q_{n}$ via $q_{m}$ and rest of initial coefficients. Finally,
the compatibility conditions $\partial _{k}(\partial _{m}q_{n})=\partial
_{k}(\partial _{m}q_{n})$ should lead to a parametrization of all
coefficients via $q_{m}$ and their derivatives with respect to moments $%
A^{0},A^{1},...,A^{M-1}$.

However, this is not precisely true. Hydrodynamic chain (\ref{link})
possesses a large class of invertible transformations, allowing to
significantly reduce a number of distinguish coefficients. For instance,
hydrodynamic chain (\ref{lin}) contains 15 coefficients, while its
integrable version (\ref{i}) contains just 3 coefficients. It means, that
transformation (\ref{mom}) is necessary but not sufficient for a most
appropriate choice of reduced number of coefficients for a satisfactory
investigation. To avoid complexity of this problem, in this Section, we
restrict our consideration on the case $M=2$ associated with hydrodynamic
chain (\ref{lin}).

\section{General solution in the \textquotedblleft
triangular\textquotedblright\ case}

In this Section, we restrict our consideration on a most important case
determined by the choice $r_{0}=1$ and $r_{1}=1$ (see (\ref{lin}) and
comments to (\ref{lina}), i.e. the restrictions $s_{0,2}=0,s_{1,2}=0$), i.e.
(see (\ref{er}))%
\begin{equation*}
\sum_{m=1}^{4}\alpha _{m}=-2.
\end{equation*}%
Moreover, we essentially can simplify further computations fixing all $\rho
_{kn}=0$, where $k,n=0,1$. Nevertheless, this is not a \textit{particular}
case. A \textit{complete} description of conservative integrable
hydrodynamic chains (\ref{tri}) is given by (\ref{i}). In general case (\ref%
{linp}), an infinite set of conservation laws can be written in the form
(cf. (\ref{tri}))%
\begin{eqnarray*}
\partial _{t}H_{k} &=&\partial _{x}F_{k}(H_{0},H_{1},...,H_{M}),\text{ }%
k=0,1,2,...,M-1, \\
\partial _{t}H_{M+k} &=&\partial _{x}F_{M+k}(H_{0},H_{1},...,H_{M+k+1}),%
\text{ }k=0,1,2,...
\end{eqnarray*}%
Just hydrodynamic chain (\ref{lin}) possesses an infinite set of
conservation laws given by (\ref{tri}). Such hydrodynamic chains we call
\textquotedblleft triangular\textquotedblright\ in comparison with all other
hydrodynamic chains (\ref{linp}), whose conservation laws possess a \textit{%
deviation} from this triangular case, i.e. first $M$ conservation law fluxes
depend simultaneously on first $M$ conservation law densities $H_{k}$.

As it was mentioned in the previous Section, the compatibility conditions $%
\partial _{k}(\partial _{q}\ln B_{1})=\partial _{q}(\partial _{k}\ln
B_{1}),\partial _{k}(\partial _{n}\ln B_{1})=\partial _{n}(\partial _{k}\ln
B_{1})$ lead to the system in involution%
\begin{equation}
\partial _{1}q_{k}=\frac{q_{k}^{2}+u_{0}q_{k}+u_{-1}}{S},\text{ \ \ \ }%
\partial _{1}S=\sum_{m=1}^{4}(2\alpha _{m}+1)q_{m},  \label{full}
\end{equation}%
\begin{equation}
\partial _{1}u_{-1}=\frac{1}{S}\left( {u_{-1}}\sum_{m=1}^{4}(\alpha
_{m}+1)q_{m}-{\underset{k=1}{\overset{4}{\prod }}q_{k}}\sum_{m=1}^{4}\frac{%
\alpha _{m}+1}{q_{m}}\right) ,  \label{zuk}
\end{equation}%
\begin{equation}
\partial _{0}q_{k}=\frac{\left( q_{k}^{2}+u_{0}q_{k}+u_{-1}\right) \left(
\partial _{0}S-u_{-1}\right) }{q_{k}S}-\frac{\partial _{0}u_{-1}}{q_{k}}%
-\partial _{0}u_{0},  \label{zik}
\end{equation}%
\begin{equation}
\partial _{0}S=-\sum_{m<k}(\alpha _{k}+\alpha _{m}+1)q_{m}q_{k},  \label{zak}
\end{equation}%
\begin{equation}
\partial _{0}u_{-1}=\frac{1}{S}\left( {\underset{k=1}{\overset{4}{\prod }}%
q_{k}}-{u_{-1}}\sum_{m<k}(\alpha _{k}+\alpha _{m}+1)q_{m}q_{k}-{(u_{-1})^{2}}%
\right) ,  \label{zyk}
\end{equation}%
where%
\begin{equation}
u_{0}=\sum_{m=1}^{4}\alpha _{m}q_{m},\text{ \ \ }S=A^{0}u_{-1}+A^{1}u_{0}-%
\sigma ,  \label{sikma}
\end{equation}%
and 6 variable coefficients are connected with 3 others by (here $k=0,1$
only) 
\begin{equation*}
\gamma _{0,k}=\partial _{k}u_{0},\text{ \ \ }\gamma _{-1,k}=\partial
_{k}u_{-1},\text{ \ \ }\omega _{k}=-\partial _{k}\sigma .
\end{equation*}%
In this case, all coefficients (\ref{he}), (\ref{hi}), (\ref{hy}), (\ref{hu}%
) significantly reduce, then hydrodynamic chain (\ref{lin}) transforms to a
more compact form given by (\ref{i}).

Let us introduce the auxiliary functions%
\begin{equation*}
\tilde{q}_{k}=\frac{q_{k}-q_{4}}{S},\text{ \ }k=1,2,3.
\end{equation*}%
Then equations (\ref{full}) reduce to the form%
\begin{eqnarray}
\partial _{1}\tilde{q}_{k} &=&\tilde{q}_{k}\left( \tilde{q}%
_{k}-\sum_{m=1}^{3}(\alpha _{m}+1)\tilde{q}_{m}\right) ,\text{ \ }k=1,2,3,
\label{a} \\
&&  \notag \\
\partial _{1}\ln S &=&\sum_{m=1}^{3}(2\alpha _{m}+1)\tilde{q}_{m},\text{ \ \ 
}u_{-1}=S\left( \partial _{1}q_{4}-q_{4}\sum_{m=1}^{3}\alpha _{m}\tilde{q}%
_{m}\right) +q_{4}^{2}.  \label{b}
\end{eqnarray}%
A substitution $u_{-1}$ from the above system into (\ref{zuk}) leads to the
simple equation of the second order%
\begin{equation}
\partial _{1}^{2}q_{4}+{(\alpha _{4}+1)}\tilde{q}_{1}\tilde{q}_{2}\tilde{q}%
_{3}S=0.  \label{secon}
\end{equation}

Let us introduce an intermediate function $z=\partial _{1}q_{4}$ and four
functions $c_{k}(A^{0}),k=0,1,2,3$.

\textbf{Lemma}: \textit{A general solution of system }(\ref{a}) \textit{is
given by}%
\begin{equation}
\tilde{q}_{k}=-\partial _{1}\ln (z-c_{k}),\text{ \ }k=1,2,3,  \label{dva}
\end{equation}%
\textit{where}%
\begin{equation}
\partial _{1}z=c_{0}{\underset{m=1}{\overset{3}{\prod }}}(z-c_{m})^{\alpha
_{m}+1},\text{ \ \ }S=\frac{c_{0}^{-2}}{{\alpha _{4}+1}}{\underset{m=1}{%
\overset{3}{\prod }}}(z-c_{m})^{-(2\alpha _{m}+1)}.  \label{muk}
\end{equation}

\textbf{Proof}: A substitution above formulas into (\ref{a}), (\ref{b}), (%
\ref{secon}) yields identities.

A substitution (\ref{dva}) and (\ref{muk}) in (\ref{zik}) and (\ref{zak})
determines%
\begin{equation*}
\partial _{0}q_{4}=\frac{c_{0}^{-1}}{{\alpha _{4}+1}}{\underset{m=1}{\overset%
{3}{\prod }}}(z-c_{m})^{-\alpha _{m}}-zq_{4},
\end{equation*}%
where $c_{0}(A^{0})$ can be found by quadratures%
\begin{equation*}
(\ln c_{0})^{\prime }=-\sum_{m=1}^{3}\left( \alpha _{m}+\alpha _{4}+1\right)
c_{m},
\end{equation*}%
while all other $c_{k}(A^{k})$ satisfy a \textit{new modification} of
well-known generalized Darboux--Halphen system (see detail in \cite{Abl}) 
\begin{eqnarray}
c_{1}^{\prime } &=&\frac{\alpha _{1}}{\alpha _{4}+1}[c_{1}\left(
c_{2}+c_{3}\right) -c_{2}c_{3}]-\frac{\alpha _{1}+\alpha _{4}+1}{\alpha
_{4}+1}c_{1}^{2},  \notag \\
c_{2}^{\prime } &=&\frac{\alpha _{2}}{\alpha _{4}+1}[c_{2}\left(
c_{1}+c_{3}\right) -c_{1}c_{3}]-\frac{\alpha _{2}+\alpha _{4}+1}{\alpha
_{4}+1}c_{2}^{2},  \label{upc} \\
c_{3}^{\prime } &=&\frac{\alpha _{3}}{\alpha _{4}+1}[c_{3}\left(
c_{1}+c_{2}\right) -c_{1}c_{2}]-\frac{\alpha _{3}+\alpha _{4}+1}{\alpha
_{4}+1}c_{3}^{2}.  \notag
\end{eqnarray}

\textbf{Remark}: A sole function%
\begin{equation}
s(A^{0})=\frac{c_{2}-c_{3}}{c_{1}-c_{3}}  \label{racio}
\end{equation}%
satisfies the so-called Schwarzian equation (see \cite{Abl}) 
\begin{equation*}
\frac{s^{\prime \prime \prime }}{s^{\prime }}-\frac{3(s^{\prime \prime })^{2}%
}{2(s^{\prime })^{2}}=\left( 2\frac{\alpha _{1}\alpha _{2}+\alpha _{3}\alpha
_{4}}{s(s-1)}-\frac{\left( \alpha _{2}+\alpha _{4}\right) \left( \alpha
_{1}+\alpha _{3}\right) }{s^{2}}-\frac{\left( \alpha _{2}+\alpha _{3}\right)
\left( \alpha _{1}+\alpha _{4}\right) }{(s-1)^{2}}\right) \frac{(s^{\prime
})^{2}}{2}.
\end{equation*}%
Then a solution of system (\ref{upc}) is given by%
\begin{equation*}
c_{1}=-\frac{1}{2}\left( \ln \frac{s^{\alpha _{1}+\alpha _{3}}s^{\prime }}{%
(s-1)^{\alpha _{1}-\alpha _{4}}}\right) ^{\prime },c_{2}=-\frac{1}{2}\left(
\ln \frac{(s-1)^{\alpha _{2}+\alpha _{3}}s^{\prime }}{s^{\alpha _{2}-\alpha
_{4}}}\right) ^{\prime },c_{3}=-\frac{1}{2}\left( \ln \left( s^{\alpha
_{1}+\alpha _{3}}(s-1){}^{\alpha _{2}+\alpha _{3}}s^{\prime }\right) \right)
^{\prime }.
\end{equation*}%
Under the simple linear transformation%
\begin{eqnarray*}
c_{1} &=&(1+\alpha _{1}+\alpha _{3})\omega _{1}-(1+\alpha _{3}+\alpha
_{4})\omega _{2}+(1-\alpha _{1}+\alpha _{4})\omega _{3}, \\
c_{2} &=&(1-\alpha _{2}+\alpha _{4})\omega _{1}-(1+\alpha _{3}+\alpha
_{4})\omega _{2}+(1+\alpha _{2}+\alpha _{3})\omega _{3}, \\
c_{3} &=&(1+\alpha _{1}+\alpha _{3})\omega _{1}+(1-\alpha _{3}+\alpha
_{4})\omega _{2}+(1+\alpha _{2}+\alpha _{3})\omega _{3},
\end{eqnarray*}%
the above formulas reduce to the form derived in \cite{Abl}, i.e.%
\begin{eqnarray*}
\omega _{1} &=&\omega _{2}\omega _{3}-\omega _{1}(\omega _{2}+\omega
_{3})+\omega ^{2}, \\
\omega _{2} &=&\omega _{1}\omega _{3}-\omega _{2}(\omega _{1}+\omega
_{3})+\omega ^{2}, \\
\omega _{3} &=&\omega _{1}\omega _{2}-\omega _{3}(\omega _{1}+\omega
_{2})+\omega ^{2},
\end{eqnarray*}%
where%
\begin{equation*}
\omega ^{2}=\beta _{1}^{2}(\omega _{1}-\omega _{2})(\omega _{3}-\omega
_{1})+\beta _{2}^{2}(\omega _{2}-\omega _{3})(\omega _{1}-\omega _{2})+\beta
_{3}^{2}(\omega _{3}-\omega _{1})(\omega _{2}-\omega _{3}),
\end{equation*}%
\begin{equation*}
\omega _{1}=-\frac{1}{2}\left( \ln \frac{s^{\prime }}{s-1}\right) ^{\prime },%
\text{ \ \ }\omega _{2}=-\frac{1}{2}\left( \ln \frac{s^{\prime }}{s(s-1)}%
\right) ^{\prime },\text{ \ \ }\omega _{3}=-\frac{1}{2}\left( \ln \frac{%
s^{\prime }}{s}\right) ^{\prime },
\end{equation*}

\begin{equation*}
\alpha _{1}=\frac{1}{2}\left( \beta _{1}+\beta _{2}-\beta _{3}-1\right)
,\alpha _{2}=\frac{1}{2}\left( -\beta _{1}+\beta _{2}+\beta _{3}-1\right)
,\alpha _{3}=\frac{1}{2}\left( \beta _{1}-\beta _{2}+\beta _{3}-1\right)
\end{equation*}%
and (\ref{racio}) reduces to%
\begin{equation*}
s(A^{0})=\frac{\omega _{2}-\omega _{1}}{\omega _{2}-\omega _{3}}.
\end{equation*}

The function $q_{4}(A^{0},z)$ can be found by the quadrature%
\begin{equation*}
dq_{4}=c_{0}^{-1}z{\underset{m=1}{\overset{3}{\prod }}}(z-c_{m})^{-\alpha
_{m}-1}dz+\frac{c_{0}^{-1}}{{\alpha _{4}+1}}{\underset{m=1}{\overset{3}{%
\prod }}}(z-c_{m})^{-\alpha _{m}-1}P_{2}(z)dA^{0},
\end{equation*}%
where $P_{2}(z)$ is a polynomial in $z$ of the second degree, i.e.%
\begin{equation*}
P_{2}(z)=z^{2}\sum_{m=1}^{3}(\alpha _{m}+\alpha _{4}+1)c_{m}+z{\underset{n=1}%
{\overset{3}{\prod }}}c_{n}\sum_{m=1}^{3}\frac{\alpha _{m}+1}{c_{m}}-{%
\underset{n=1}{\overset{3}{\prod }}}c_{n}.
\end{equation*}%
Then the first moment $A^{1}(A^{0},z)$ is determined by another quadrature

\begin{equation*}
dA^{1}=c_{0}^{-1}{\underset{m=1}{\overset{3}{\prod }}}(z-c_{m})^{-\alpha
_{m}-1}dz+\left( \frac{c_{0}^{-1}}{{\alpha _{4}+1}}{\underset{m=1}{\overset{3%
}{\prod }}}(z-c_{m})^{-\alpha _{m}-1}G_{2}(z)+q_{4}\right) dA^{0},
\end{equation*}%
where $G_{2}(z)$ is a polynomial in $z$ of the second degree, i.e.%
\begin{equation*}
G_{2}(z)=-z^{2}+z\sum_{m=1}^{3}(\alpha _{m}+\alpha _{4}+2)c_{m}+{\underset{%
n=1}{\overset{3}{\prod }}}c_{n}\sum_{m=1}^{3}\frac{\alpha _{m}}{c_{m}}.
\end{equation*}%
Thus, all functions $u_{0},u_{-1}$ and $\sigma $ (see (\ref{sikma}), the
second formula in (\ref{b}) and the second formula in (\ref{muk})) are
expressed via above implicit dependencies $z(A^{0},A^{1}),q_{4}(A^{0},A^{1})$%
, i.e.%
\begin{equation*}
\sigma =A^{0}u_{-1}+A^{1}u_{0}-\frac{c_{0}^{-2}}{{\alpha _{4}+1}}{\underset{%
m=1}{\overset{3}{\prod }}}(z-c_{m})^{-(2\alpha _{m}+1)},
\end{equation*}%
\begin{equation*}
u_{0}=-\frac{1}{\left( \alpha _{4}+1\right) c_{0}}\sum_{m=1}^{3}\frac{\alpha
_{m}}{z-c_{m}}\prod_{k=1}^{3}\left( z-c_{k}\right) {}^{-\alpha _{k}}-2q_{4},
\end{equation*}%
\begin{equation*}
u_{-1}=\frac{z}{{}\left( \alpha _{4}+1\right) c_{0}^{2}}\prod_{k=1}^{3}%
\left( z-c_{k}\right) {}^{-\left( 2\alpha _{k}+1\right) }+\frac{q_{4}}{%
\left( \alpha _{4}+1\right) c_{0}}\sum_{m=1}^{3}\frac{\alpha _{m}}{z-c_{m}}%
\prod_{k=1}^{3}\left( z-c_{k}\right) {}^{-\alpha _{k}}+q_{4}^{2}.
\end{equation*}

\section{Conservative hydrodynamic chains}

We believe that integrable hydrodynamic chain (\ref{i}) must possess an
infinite set of conservation laws (\ref{tri}). It means, that an infinite
series of invertible triangular transformations $%
H_{k}(A^{0},A^{1},...,A^{k}) $ can be found. In this Section, we present a
generating function of conservation laws and prove an equivalence of the
system in involution derived by E.V. Ferapontov and D.G. Marshall for
conservative hydrodynamic chains (\ref{tri}) with system in involution (\ref%
{full})--(\ref{zyk}) derived in the previous Section.

\textbf{Theorem}: \textit{Integrable hydrodynamic chain} (\ref{i}) \textit{%
possesses a generating function of conservation laws}%
\begin{equation}
\partial _{t}p(q,A^{0},A^{1})=\partial _{x}Q(q,A^{0},A^{1}).  \label{ken}
\end{equation}

\textbf{Proof}: Integrable hydrodynamic chain (\ref{i}) is associated with
the Vlasov type kinetic equation (see (\ref{linp}))%
\begin{equation*}
q_{t}=qq_{x}-\left( q^{2}+u_{0}q+u_{-1}\right) [\log (A^{2}+\sigma
)]_{x}+q(u_{0})_{x}+(u_{-1})_{x}.
\end{equation*}%
First two equations of hydrodynamic chain (\ref{i}) are given by%
\begin{equation*}
A_{t}^{0}=A_{x}^{1},\text{ \ \ }%
A_{t}^{1}=A_{x}^{2}-[(A^{2}+u_{0}A^{1}+u_{-1}A^{0})[\ln (A^{2}+\sigma
)]_{x}-A^{1}(u_{0})_{x}-A^{0}(u_{-1})_{x}].
\end{equation*}%
Since r.h.s. of $q_{t}$ and $A_{t}^{1}$ contain derivative $A_{x}^{2}$,
formally $Q$ should depend on $A^{2}$. Differentiation (\ref{ken}) with
respect to $x$ and $t$%
\begin{equation*}
\partial _{q}p\cdot q_{t}+\partial _{0}p\cdot A_{t}^{0}+\partial _{1}p\cdot
A_{t}^{1}=\partial _{q}Q\cdot q_{x}+\partial _{0}Q\cdot A_{x}^{0}+\partial
_{1}Q\cdot A_{x}^{1}+\partial _{2}Q\cdot A_{x}^{2}
\end{equation*}%
leads to\footnote{%
as usual, we are looking for a general solution. It means, that all $%
A_{x}^{k}$ are considered independently. Thus, corresponding coefficients
must vanish separately.}%
\begin{eqnarray*}
\partial _{q}Q &=&q\partial _{q}p, \\
\partial _{0}Q &=&\left( \partial _{0}u_{-1}+q\partial _{0}u_{0}-\frac{%
(q^{2}+qu_{0}+u_{-1})\partial _{0}\sigma }{\sigma +A_{2}}\right) \partial
_{q}p \\
&+&\left( A_{0}\partial _{0}u_{-1}+A_{1}\partial _{0}u_{0}-\frac{\left(
A_{2}+A_{0}u_{-1}+A_{1}u_{0}\right) \partial _{0}\sigma }{\sigma +A_{2}}%
\right) \partial _{1}p, \\
\partial _{1}Q &=&\partial _{0}p+\left( \partial _{1}u_{-1}+q\partial
_{1}u_{0}-\frac{(q^{2}+qu_{0}+u_{-1})\partial _{1}\sigma }{\sigma +A_{2}}%
\right) \partial _{q}p \\
&+&\left( A_{0}\partial _{1}u_{-1}+A_{1}\partial _{1}u_{0}-\frac{\left(
A_{2}+A_{0}u_{-1}+A_{1}u_{0}\right) \partial _{1}\sigma }{\sigma +A_{2}}%
\right) \partial _{1}p, \\
\partial _{2}Q &=&\frac{\sigma -A_{0}u_{-1}-A_{1}u_{0}}{\sigma +A_{2}}%
\partial _{1}p-\frac{q^{2}+qu_{0}+u_{-1}}{\sigma +A_{2}}\partial _{q}p.
\end{eqnarray*}%
If $\partial _{2}Q=0$, then%
\begin{equation}
\partial _{1}p=\frac{q^{2}+qu_{0}+u_{-1}}{\sigma -A_{0}u_{-1}-A_{1}u_{0}}%
\partial _{q}p  \label{svya}
\end{equation}%
and all other above expressions simplify to the form 
\begin{eqnarray*}
\partial _{q}Q &=&q\partial _{q}p,\text{ \ }\partial _{0}Q=\left( \frac{%
\left( q^{2}+qu_{0}+u_{-1}\right) \left( u_{-1}-\partial _{0}S\right) }{S}%
+\partial _{0}u_{-1}+q\partial _{0}u_{0}\right) \partial _{q}p, \\
\partial _{1}Q &=&\partial _{0}p+\left( \frac{\left(
q^{2}+qu_{0}+u_{-1}\right) \left( u_{0}-\partial _{1}S\right) }{S}+\partial
_{1}u_{-1}+q\partial _{1}u_{0}\right) \partial _{q}p.
\end{eqnarray*}%
A compatibility conditions $\partial _{1}(\partial _{0}Q)=\partial
_{0}(\partial _{1}Q),\partial _{1}(\partial _{q}Q)=\partial _{q}(\partial
_{1}Q),\partial _{q}(\partial _{0}Q)=\partial _{0}(\partial _{q}Q)$ lead to
three equations containing four second order derivatives $\partial
_{0q}p,\partial _{qq}p,\partial _{1q}p,\partial _{00}p$ only. Taking into
account (\ref{svya}), the derivative $\partial _{1q}p$ is proportional to $%
\partial _{qq}p$, and all other three derivatives $\partial _{0q}p,\partial
_{qq}p,\partial _{00}p$ can be expressed. Moreover, a direct further
computation leads to the correspondence%
\begin{equation}
\partial _{q}p=\frac{1}{B_{1}}.  \label{z}
\end{equation}%
Thus, (see (\ref{ba}))%
\begin{equation}
\partial _{q}p=\overset{4}{\underset{m=1}{\prod }}(q-q_{m})^{\alpha _{m}}.
\label{expand}
\end{equation}%
The generating function of conservation law densities can be found in two
quadratures (see (\ref{svya}) and (\ref{z}))%
\begin{equation*}
dp=\overset{4}{\underset{m=1}{\prod }}(q-q_{m})^{-\alpha _{m}}dq-\frac{%
q^{2}+qu_{0}+u_{-1}}{S}\overset{4}{\underset{m=1}{\prod }}(q-q_{m})^{-\alpha
_{m}}dA^{1}+(\partial _{0}p)dA^{0},
\end{equation*}%
where $\partial _{0}p$ also is determined by corresponding second derivatives%
\begin{equation*}
d(\partial _{0}p)=(\partial _{0q}p)dq+(\partial _{01}p)dA^{1}+(\partial
_{00}p)dA^{0}.
\end{equation*}%
Nevertheless, an infinite series of conservation law densities can be found
directly from (\ref{expand}).

In contrary with the above approach, all conservation laws can be found
iteratively. The zeroth conservation law is given by the zeroth equation%
\begin{equation*}
\partial _{t}H_{0}=\partial _{x}F_{0}(H_{0},H_{1}),
\end{equation*}%
such that $A^{0}=H_{0}$ and $A^{1}=F_{0}(H_{0},H_{1})$ (see integrable
hydrodynamic chain (\ref{i})). Let us introduce an intermediate notation $%
h=\partial _{0}H_{1}$.

\textbf{Lemma}: \textit{Integrable hydrodynamic chain} (\ref{i}) \textit{%
possesses first conservation law}%
\begin{equation*}
\partial _{t}H_{1}=\partial _{x}[\ln H_{2}+G(H_{0},H_{1})],
\end{equation*}%
\textit{such that the second conservation law density}%
\begin{equation*}
H_{2}=\frac{1}{A^{2}+\sigma },
\end{equation*}%
\textit{the first conservation law density} $H_{1}$ \textit{can be found by
two quadratures}%
\begin{equation}
dH_{1}=hdA^{0}+\frac{1}{S}dA^{1},\text{ \ }dh=\left( \partial _{0}\frac{u_{0}%
}{S}-\partial _{1}\frac{u_{-1}}{S}\right) dA^{0}+\left( \partial _{0}\frac{1%
}{S}\right) dA^{1},  \label{e}
\end{equation}%
\textit{and the function} $G(H_{0},H_{1})$ \textit{is determined by the
quadrature} 
\begin{equation*}
dG=\frac{\partial _{0}S-u_{-1}}{S}dA^{0}+\left( h+\frac{\partial _{1}S-u_{0}%
}{S}\right) dA^{1}.
\end{equation*}

\textbf{Proof}: can be obtained by a straightforward computation.

The system in involution on third derivatives of functions $F(H_{0},H_{1})$
and $G(H_{0},H_{1})$ was derived (see (\ref{ferma}) in the Appendix) in
paper \cite{FerMarsh}.

\textbf{Theorem}: \textit{These functions} $F(H_{0},H_{1})$ \textit{and} $%
G(H_{0},H_{1})$ \textit{can be found in quadratures}%
\begin{eqnarray*}
dG &=&\left( \frac{\tilde{\partial}_{0}S-u_{-1}}{S}+u_{0}h-Sh^{2}\right)
dH_{0}+\left( \tilde{\partial}_{1}\ln S-u_{0}+hS\right) dH_{1}, \\
dF &=&-hSdH_{0}+SdH_{1},
\end{eqnarray*}%
\textit{where} $\tilde{\partial}_{0}\equiv \partial _{H_{0}},\tilde{\partial}%
_{1}\equiv \partial _{H_{1}}$. \textit{An inverse transformation is given by}

\begin{equation*}
u_{0}=\frac{\tilde{\partial}_{1,1}F}{\tilde{\partial}_{1}F}-\tilde{\partial}%
_{0}F-\tilde{\partial}_{1}G,\text{ \ }S=\tilde{\partial}_{1}F,\text{ \ }%
u_{-1}=\tilde{\partial}_{0,1}F-\frac{\tilde{\partial}_{0}F\tilde{\partial}%
_{1,1}F}{\tilde{\partial}_{1}F}+\tilde{\partial}_{0}F\tilde{\partial}_{1}G-%
\tilde{\partial}_{1}F\tilde{\partial}_{0}G.
\end{equation*}

\textbf{Proof}: can be obtained by a straightforward computation.

\textbf{Remark}: All higher commuting flows belong to (\ref{lina}) in a
general case. Indeed, a first commuting flow to hydrodynamic chain (\ref{i})
is given by%
\begin{equation*}
A_{y}^{k}=\left[ \left( A^{k+2}+A^{k+1}u_{0}+A^{k}u_{-1}\right) H_{2}\right]
_{x}+k\Big(\left( A^{k+2}+A^{k+1}u_{0}+A^{k}u_{-1}\right) (H_{2})_{x}-\Big.
\end{equation*}%
\begin{equation*}
\left( A^{k+1}+A^{k}u_{0}+A^{k-1}u_{-1}\right) \left( \frac{H_{3}}{H_{2}}-%
\frac{H_{2}u_{0}}{2}\right) _{x}
\end{equation*}%
\begin{equation*}
+\Big.\frac{1}{2}\left[ A^{k}\left( u_{0,H_{1}}\partial _{x}+\partial
_{x}u_{0,H_{1}}\right) +A^{k-1}\left( u_{-1,H_{1}}\partial _{x}+\partial
_{x}u_{-1,H_{1}}\right) \right] H_{2}\Big)
\end{equation*}%
where the third conservation law density is determined by (here $%
u_{0,H_{1}}=\partial _{H_{1}}u_{0},u_{-1,H_{1}}=\partial
_{H_{1}}u_{-1},\sigma _{H_{1}}=\partial _{H_{1}}\sigma $)%
\begin{equation*}
H_{3}=(A^{3}-\sigma u_{0}+A^{1}u_{-1}+\frac{1}{2}\sigma _{H_{1}})H_{2}^{3}+%
\frac{3}{2}u_{0}H_{2}^{2}.
\end{equation*}%
A compatibility condition $(\lambda _{t})_{y}=(\lambda _{y})_{t}$ of
corresponding Vlasov type kinetic equations (\ref{linn}) (i.e. $K=1$ and $K=2
$, respectively) leads to some 2+1 dimensional quasilinear equation of the
second order (a general classification was presented in \cite{BFT}), which
will be considered in a separate paper.

\section{Egorov's case}

A most important and interesting case is the Egorov hydrodynamic chain (see 
\cite{egor}) selected by the simple choice $H_{1}=A^{1}$ (see (\ref{e})). In
such a case, $S=1$, then all $\alpha _{k}=-1/2$ and general hydrodynamic
chain (\ref{i}) reduces to the form%
\begin{equation*}
A_{t}^{k}=A_{x}^{k+1}-k[(A^{k+1}+A^{k}\partial _{1}F+A^{k-1}\partial
_{0}F)[\ln (A^{2}+A^{1}\partial _{1}F+A^{0}\partial
_{0}F-1)]_{x}-A^{k}(\partial _{1}F)_{x}-A^{k-1}(\partial _{0}F)_{x}],
\end{equation*}%
where the function $F$ is given by%
\begin{equation*}
F=\frac{1}{4}\int \eta (A^{0})dA^{0}+\ln \theta _{1}(A^{1},A^{0}).
\end{equation*}%
Here $\eta (A^{0})$ is a solution of the Chazy equation%
\begin{equation*}
\eta ^{\prime \prime \prime }=3\eta ^{\prime ^{2}}-2\eta \eta ^{\prime
\prime }
\end{equation*}%
and the Jacobi theta-function%
\begin{equation*}
\theta _{1}(A^{1},A^{0})=2\sum_{n=0}^{\infty
}(-1)^{n}e^{-(n+1/2)^{2}A^{0}}\sin [(2n+1)A^{1}]
\end{equation*}%
is connected with the above solution of the Chazy equation via an involutive
system (see \cite{egor})%
\begin{eqnarray*}
\partial _{1}\theta _{1} &=&-\mu \theta _{1},\text{ \ \ }\partial _{0}\theta
_{1}=\frac{1}{4}(\mu ^{2}-l)\theta _{1}, \\
\partial _{1}\mu  &=&l,\text{ \ \ \ }\partial _{0}\mu =\frac{1}{4}\sqrt{%
4l^{3}-4\eta l^{2}-8\eta ^{\prime }l-\frac{8}{3}\eta ^{\prime \prime }}-%
\frac{1}{2}\mu l, \\
\partial _{1}l &=&\sqrt{4l^{3}-4\eta l^{2}-8\eta ^{\prime }l-\frac{8}{3}\eta
^{\prime \prime }},\text{ \ }\partial _{0}l=l^{2}-\eta l-\eta ^{\prime }-%
\frac{1}{2}\mu \sqrt{4l^{3}-4\eta l^{2}-8\eta ^{\prime }l-\frac{8}{3}\eta
^{\prime \prime }}.
\end{eqnarray*}

\section{Conclusion}

The crucial observation made in \cite{MaksHam} is that a substitution of
Zakharov moment decomposition (\ref{zakh}) is applicable for hydrodynamic
chains, whose r.h.s. expressions depend \textit{linearly} on a discrete
variable $k$ and contain a finite number of common variable coefficients
(see (\ref{lina})).

In comparison with approaches established earlier (see \cite{FK}, \cite%
{FerMarsh}, \cite{GT}), the method presented in this paper is not universal
but most effective. A complete classification of conservative hydrodynamic
chains is given by virtue of their re-presentation in a special form (\ref{5}%
). All conservation law densities $H_{m}$ can be expressed explicitly via
moments $A^{k}$; all fluxes of corresponding conservation laws can be
expressed explicitly via $H_{m}$; all commuting flows can be constructed
explicitly (and their conservation laws); infinitely many hydrodynamic
reductions can be extracted. Thus, infinitely many particular solutions to
integrable hydrodynamic chains (\ref{5}) can be presented (by the
generalized hodograph method, see \cite{Tsar}).

\section{Appendix}

The system in involution for functions $u_{0}$, $u_{-1}$ and $S$ describing
a family of integrable hydrodynamic chains (\ref{i}) possesses a general
solution parameterized by 9 arbitrary constants:%
\begin{eqnarray*}
\partial _{1,1}{}u_{-1} &=&\partial _{1}u_{-1}\cdot \partial _{1}\ln S+2%
\frac{\partial _{0}u_{-1}-\text{ }u_{-1}\partial _{1}u_{0}}{S}, \\
\partial _{1,1}{}u_{0} &=&\partial _{1}u_{0}\cdot \partial _{1}\ln S+2\frac{%
\partial _{1}u_{-1}+\partial _{0}u_{0}-u_{0}\partial _{1}u_{0}}{S}, \\
\partial _{0,1}u_{-1} &=&\partial _{1}u_{-1}\cdot \partial _{0}\ln S-2\frac{%
u_{-1}\left( \partial _{1}u_{-1}+\partial _{0}u_{0}\right) -u_{0}\text{ }%
\partial _{0}u_{-1}}{S}, \\
\partial _{0,1}u_{0} &=&\partial _{1}u_{0}\cdot \partial _{0}\ln S+2\frac{%
\partial _{0}u_{-1}-u_{-1}\partial _{1}u_{0}}{S}, \\
\partial _{0,1}S &=&\partial _{1}u_{-1}-\partial _{0}u_{0}+\frac{\partial
_{0}S-2u_{-1}}{S}\partial _{1}S+\frac{u_{0}\partial _{0}S}{S},
\end{eqnarray*}%
\begin{equation}
\partial _{1,1}{}{}S=\frac{\partial _{1}^{2}S{}}{S}-\frac{u_{0}\partial _{1}S%
}{S}+\frac{2\partial _{0}S}{S},  \label{usi}
\end{equation}%
\begin{eqnarray*}
\partial _{0,0}{}S &=&\frac{2\partial _{0,0}S{}}{S}+\frac{\left(
u_{0}^{2}-2u_{-1}\right) \partial _{0}S}{S}+u_{-1}\frac{\partial _{1,1}S{}}{S%
}+\left( \partial _{1}u_{-1}-\partial _{0}u_{0}\right) u_{0} \\
&&+\left( \partial _{0}u_{0}-\partial _{1}u_{-1}-\frac{u_{0}}{S}\partial
_{0}S-\frac{u_{-1}u_{0}}{S}\right) \partial _{1}S, \\
\partial _{0,0}{}u_{0} &=&\left( \partial _{0}u_{0}-\partial
_{1}u_{-1}\right) \partial _{1}u_{0}+\frac{u_{-1}\partial
_{1}u_{0}-2\partial _{0}u_{-1}}{S}\partial _{1}S \\
&&-2\frac{u_{-1}\left( \partial _{1}u_{-1}+\partial _{0}u_{0}\right)
-u_{0}\partial _{0}u_{-1}}{S}+\frac{2\left( \partial _{1}u_{-1}+\partial
_{0}u_{0}\right) -u_{0}\partial _{1}u_{0}}{S}\partial _{0}S, \\
\partial _{0,0}{}u_{-1} &=&2\partial _{0}u_{-1}\cdot \partial
_{1}u_{0}-\left( \partial _{1}u_{-1}+\partial _{0}u_{0}\right) \partial
_{1}u_{-1}+\frac{u_{-1}\left( \partial _{1}u_{-1}+2\partial _{0}u_{0}\right)
-2\text{ }u_{0}\partial _{0}u_{-1}}{S}\partial _{1}S
\end{eqnarray*}%
\begin{equation*}
+\frac{2(\partial _{0}u_{-1}-u_{-1}\partial _{1}u_{0})+u_{0}\partial
_{1}u_{-1}}{S}\partial _{0}S+2\frac{u_{-1}^{2}\partial
_{1}u_{0}-u_{-1}\left( \partial _{0}u_{-1}+\left( \partial
_{1}u_{-1}+\partial _{0}u_{0}\right) u_{0}\right) +u_{0}^{2}\partial
_{0}u_{-1}}{S}.
\end{equation*}

The system in involution describing conservative hydrodynamic chains (\ref%
{tri}) was derived in \cite{FerMarsh}:%
\begin{equation*}
\tilde{\partial}_{0,0,0}G=\frac{2\tilde{\partial}_{0,1}^{2}G}{\tilde{\partial%
}_{1}F_{0}}+\frac{\left( -4\tilde{\partial}_{0}G\cdot \tilde{\partial}_{1}G+4%
\tilde{\partial}_{0}G\cdot \tilde{\partial}_{0}F_{0}-\tilde{\partial}%
_{0}^{2}F_{0}\right) \tilde{\partial}_{0,1}G}{\tilde{\partial}_{1}F_{0}}
\end{equation*}%
\begin{equation*}
+\left( \frac{2\tilde{\partial}_{0}^{2}G}{\tilde{\partial}_{1}F_{0}}-\frac{2%
\tilde{\partial}_{0,0}G}{\tilde{\partial}_{1}F_{0}}\right) \tilde{\partial}%
_{1,1}G-\frac{\left( \left( \tilde{\partial}_{0}F_{0}-\tilde{\partial}%
_{1}G\right) \tilde{\partial}_{0,0}F_{0}+2\tilde{\partial}_{0}G\cdot \tilde{%
\partial}_{0,1}F_{0}\right) \tilde{\partial}_{0}G}{\tilde{\partial}_{1}F_{0}}
\end{equation*}

\begin{equation*}
+\left( 2\tilde{\partial}_{0}G+\frac{2\left( \tilde{\partial}_{1}^{2}G-2%
\tilde{\partial}_{0}F_{0}\cdot \tilde{\partial}_{1}G+\tilde{\partial}%
_{0}^{2}F_{0}{}+\tilde{\partial}_{0,1}F_{0}\right) }{\tilde{\partial}%
_{1}F_{0}}\right) \tilde{\partial}_{0,0}G,
\end{equation*}%
\begin{equation*}
\tilde{\partial}_{1,1,1}G=-\frac{\left( \tilde{\partial}_{1}G-\tilde{\partial%
}_{0}F_{0}\right) {}^{2}\tilde{\partial}_{1,1}F_{0}}{\tilde{\partial}%
_{1}F_{0}}+4\tilde{\partial}_{0,1}G\cdot \tilde{\partial}_{1}F_{0}-\tilde{%
\partial}_{0,0}F_{0}\cdot \tilde{\partial}_{1}F_{0}
\end{equation*}%
\begin{equation*}
-2\tilde{\partial}_{1}G\cdot \tilde{\partial}_{0,1}F_{0}+2\tilde{\partial}%
_{0}F_{0}\cdot \tilde{\partial}_{0,1}F_{0}-\tilde{\partial}_{0}G\cdot \tilde{%
\partial}_{1,1}F_{0}+\left( 2\left( \tilde{\partial}_{1}G-\tilde{\partial}%
_{0}F_{0}\right) +\frac{\tilde{\partial}_{1,1}F_{0}}{\tilde{\partial}%
_{1}F_{0}}\right) \tilde{\partial}_{1,1}G,
\end{equation*}%
\begin{equation}
\tilde{\partial}_{0,1,1}G=2\tilde{\partial}_{0}G\cdot \tilde{\partial}%
_{1,1}G+2\tilde{\partial}_{0,0}G\cdot \tilde{\partial}_{1}F_{0}-2\tilde{%
\partial}_{0}G\cdot \tilde{\partial}_{0,1}F_{0}  \label{ferma}
\end{equation}%
\begin{equation*}
+\frac{\tilde{\partial}_{0,1}G\cdot \tilde{\partial}_{1,1}F_{0}}{\tilde{%
\partial}_{1}F_{0}}+\frac{\left( \tilde{\partial}_{0}F_{0}-\tilde{\partial}%
_{1}G\right) \tilde{\partial}_{0}G\cdot \tilde{\partial}_{1,1}F_{0}}{\tilde{%
\partial}_{1}F_{0}},
\end{equation*}%
\begin{equation*}
\tilde{\partial}_{0,0,1}G=-\tilde{\partial}_{0}^{2}G{}\frac{\tilde{\partial}%
_{1,1}F_{0}}{\tilde{\partial}_{1}F_{0}}+4\tilde{\partial}_{0,1}G\cdot \tilde{%
\partial}_{0}G-\tilde{\partial}_{0,0}F_{0}\cdot \tilde{\partial}_{0}G+\left(
2\tilde{\partial}_{0}F_{0}-2\tilde{\partial}_{1}G+\frac{\tilde{\partial}%
_{1,1}F_{0}}{\tilde{\partial}_{1}F_{0}}\right) \tilde{\partial}_{0,0}G,
\end{equation*}

\begin{equation*}
\tilde{\partial}_{1,1,1}F_{0}=\frac{\tilde{\partial}_{1,1}^{2}F_{0}}{\tilde{%
\partial}_{1}F_{0}}+\left( \tilde{\partial}_{1}G-\tilde{\partial}%
_{0}F_{0}\right) \tilde{\partial}_{1,1}F_{0}+2\tilde{\partial}_{1}F_{0}\cdot 
\tilde{\partial}_{0,1}F_{0},
\end{equation*}%
\begin{equation*}
\tilde{\partial}_{0,1,1}F_{0}=\tilde{\partial}_{0,0}F_{0}\cdot \tilde{%
\partial}_{1}F_{0}+\left( \tilde{\partial}_{0}G+\frac{\tilde{\partial}%
_{0,1}F_{0}}{\tilde{\partial}_{1}F_{0}}\right) \tilde{\partial}_{1,1}F_{0},
\end{equation*}%
\begin{equation*}
\tilde{\partial}_{0,0,1}F_{0}=2\tilde{\partial}_{0}G\cdot \tilde{\partial}%
_{0,1}F_{0}+\left( \tilde{\partial}_{0}F_{0}-\tilde{\partial}_{1}G+\frac{%
\tilde{\partial}_{1,1}F_{0}}{\tilde{\partial}_{1}F_{0}}\right) \tilde{%
\partial}_{0,0}F_{0},
\end{equation*}%
\begin{equation*}
\tilde{\partial}_{0,0,0}F_{0}=\left( \tilde{\partial}_{0}G+\frac{\tilde{%
\partial}_{1}^{2}G{}-2\tilde{\partial}_{0}F_{0}\cdot \tilde{\partial}_{1}G+%
\tilde{\partial}_{0}^{2}F_{0}{}-\tilde{\partial}_{1,1}G}{\tilde{\partial}%
_{1}F_{0}}\right) \tilde{\partial}_{0,0}F_{0}+
\end{equation*}%
\begin{equation*}
\left( \frac{\tilde{\partial}_{0,0}F_{0}}{\tilde{\partial}_{1}F_{0}}-2\frac{%
\tilde{\partial}_{0}G\cdot \tilde{\partial}_{1}G-\tilde{\partial}_{0,1}G-%
\tilde{\partial}_{0}G\cdot \tilde{\partial}_{0}F_{0}}{\tilde{\partial}%
_{1}F_{0}}\right) \tilde{\partial}_{0,1}F_{0}+\frac{\left( \tilde{\partial}%
_{0}^{2}G{}-\tilde{\partial}_{0,0}G\right) \tilde{\partial}_{1,1}F_{0}}{%
\tilde{\partial}_{1}F_{0}}.
\end{equation*}

\section*{Acknowledgement}

Authors thank Vsevolod Adler, Boris Dubrovin, Eugeni Ferapontov, Igor
Krichever, Vladimir Sokolov and Sergey Tsarev for their stimulating and
clarifying discussions.

MVP is grateful to the SISSA in Trieste (Italy) where some part of this work
has been done. This research was particularly supported by the RFBR grant
08-01-00464) and by the grant of Presidium of RAS \textquotedblleft
Fundamental Problems of Nonlinear Dynamics\textquotedblright .

\end{document}